\documentstyle[jkas]{article}

\beginpage{1}
\endpage{?}
\year{2011} 
\volume{?} 
\month{December}
\runningauthor{J. Kim et al.} 
\runningtitle{Cosmological $N$-body simulations} 
\month{December} 
\year{2011} 
\volume{44} 
\issueno{6}
\beginpage{217}
\endpage{234}
\date{Received November 9, 2011; Revised November 22, 2011; Accepted December 2, 2011}

\begin{document}



\title{The New Horizon Run Cosmological $N$-Body Simulations}
\author{Juhan Kim$^{1}$, Changbom Park$^{2}$, Graziano Rossi$^{2}$,
  Sang Min Lee$^{3}$ and J. Richard Gott III$^{4}$} 
\address{$^1$ Center for Advanced Computation, Korea Institute for
  Advanced Study, 85 Hoegiro, Dongdaemun-gu, Seoul 130-722, Korea\\
 {\it E-mail : kjhan@kias.re.kr}}
\address{$^2$ School of Physics, Korea Institute for Advanced Study, 85 Hoegiro, Dongdaemun-gu, Seoul 130-722, Korea}
\address{$^3$ Supercomputing Center, KISTI, 335 Gwahangno, Yuseong-gu,
  Daejon, 305-806 Korea}
\address{$^4$ Department of Astrophysical Sciences, Princeton
  University, Princeton, NJ 08550, USA}
\address{\normalsize{\it (November 9, 2011; Revised November 22, 2011; Accepted December 2, 2011)}}
\offprints{Graziano Rossi (graziano@kias.re.kr)}



\abstract{We present two large cosmological $N$-body
  simulations, called Horizon Run 2 (HR2) and Horizon Run 3 (HR3), made
  using $6000^3$ = 216 billions and  $7210^3$ = 374 billion
  particles, spanning a
  volume of (7.200 $h^{-1}\textrm{Gpc}$)$^3$ and (10.815 $h^{-1}\textrm{Gpc}$)$^3$,
  respectively. These simulations improve on our previous
  Horizon Run 1 (HR1) up to a factor of 4.4 in volume, and range from 2600 to
  over 8800 times the volume of
  the Millennium Run. In addition, they achieve a considerably
  finer mass resolution, down to $1.25 \times 10^{11} h^{-1} $M$_{\odot}$, allowing to resolve galaxy-size
  halos with mean particle separations of 1.2$h^{-1}\textrm{Mpc}$ and
  1.5$h^{-1}$Mpc, respectively. 
  We have measured the power spectrum, correlation function, mass
  function and basic halo properties with percent
  level accuracy, and verified that they correctly reproduce the $\Lambda\textrm{CDM}$
  theoretical expectations, in excellent agreement with linear
  perturbation theory. Our unprecedentedly large-volume $N$-body
  simulations can be used for a variety of
  studies in cosmology and astrophysics, ranging from large-scale structure
  topology, baryon acoustic oscillations, dark energy and the
  characterization of the expansion history of the
  Universe, till galaxy formation science -- in connection with the
  new SDSS-III. 
  To this end, we made a total of 35 all-sky mock surveys along the past
  light cone out to $z=0.7$ (8 from the HR2 and 27 from the HR3), to simulate the BOSS geometry.
  The simulations and mock surveys are already publicly available at \textrm{http://astro.kias.re.kr/Horizon-Run23/.}}



\maketitle



\section{INTRODUCTION}   \label{sec_intro}


State-of-the-art observations from the Cosmic Microwave
Background (CMB), such as data provided by the Wilkinson Microwave
Anisotropy Probe 
(WMAP; Spergel et al. 2003, Komatsu et al. 2011), and from the
Large Scale Structure (LSS) as in the Sloan Digital Sky
Survey (SDSS; York et al. 2000), in the 2dF galaxy redshift survey (Colless et al. 2001) and
in the WiggleZ survey (Blake et al. 2008), support a model of the Universe dominated by Cold Dark
Matter (CDM) and Dark Energy (DE), with baryons constituting only a percent
of the total matter-energy content. 
Structures form and grow hierarchically, from
the smallest objects to the largest ones.
In this framework, dark matter collapses first into small haloes
which accrete matter and merge to form progressively larger halos over
time. Galaxies form subsequently, out of gas which cools and
collapses to the centers of dark matter halos (White \& Rees 1978;
Peebles 1982; Davis et al. 1985); therefore, halo and galaxy properties are
strongly related.
In synergy with independent data from Type I supernovae (SNI$a$; Riess et al. 1998,
Perlmutter et al. 1999, Kowalski et al. 2008), we also have increasing  
confirmations that the Universe is geometrically flat, and  
currently undergoing a phase of accelerated expansion.


What is driving this expansion, and in particular the nature of dark
energy, still remains to be explained. To date, it is perhaps one of
the most important open questions in cosmology, which would deeply
impact our understanding of fundamental physics (see for example Albrecht et al. 2006). 
In addition, investigating the nature of the (nearly) Gaussian primordial fluctuations
-- which eventually led to the
formation of halos and galaxies -- is also of utmost importance for
shedding light into the structure formation process.
In fact, constraining possible deviations from Gaussianity will impact
for example the LSS topology (Gott et al. 1986; Park et al. 1998, 2005; Gott et
al. 2009; Choi et. al 2010), as any primordial non-Gaussianity might modify the clustering
properties of massive cosmic structures forming out of rare density
fluctuations (LoVerde et al. 2008; Desjacques et al. 2009; Jeong \& Komatsu 2009), and generate a scale-dependent large-scale bias in the
clustering properties of massive dark matter halos (Dalal et al. 2008;
Verde \& Matarrese 2009; Desjacques \& Seljak 2010). 
Moreover, studying global halo properties such as halo formation, density profiles,
concentrations, shapes, kinematics, assembly times, spin, velocity distributions,
substructures and the effects of baryons is particularly important in order
to gain insights into the
formation and evolution of galaxies, their large-scale environment in
the cosmic web (Bett et al. 2007;  Gao \& White 2007; Jing Suto \& Mo
2007; Li et al. 2008; Gott et al. 2009; Shandarin et al. 2010), and to test the validity of cosmological models. 


Therefore, at the moment there is considerable interest in
constraining cosmological models (i.e. the main
cosmological parameters) from real datasets with percent level
accuracy, and especially the dark energy equation of state (DE EoS).
In particular, most of the efforts towards characterizing
DE and its properties are focused on measurements of 
the baryon acoustic oscillation (BAO) scale. 
The BAO distinct signature, imprinted in the large-scale
galaxy distribution by acoustic fluctuations in the baryon-photon
fluid prior to recombination,
appears as a quasi-harmonic
series of oscillations of decreasing amplitude in the galaxy power spectrum
at wavenumbers $0.01h$Mpc$^{-1} \le k \le 0.4 h $Mpc$^{-1}$ (Sugiyama 1995;
Eisenstein \& Hu 1998, 1999), or as a broad and quasi-Gaussian peak
in the corresponding two-point correlation function (Matsubara 2004).
The measurement of its scale is often achieved by using a well-controlled sample of 
luminous red galaxies (LRGs), as observed for example in the SDSS or
in the 2dFGRS surveys (Colless et al. 2001; Eisenstein et al. 2005; Cole et
al. 2005; Sanchez et al 2006, 2009; Percival et al. 2007, 2010;
Gaztanaga et al. 2009; Kazin et al. 2010; Reid et
al. 2010; Carnero et al. 2011).
Since the BAO scale depends solely on the plasma physics after the
Big Bang and  can be calibrated using CMB data, it can be used as a
standard ruler to measure the redshift dependence of the Hubble parameter
$H(z)$ and the angular diameter distance (Montesano et al. 2010), and thus to constrain the DE EoS parameter
$w(z)$ as a function of redshift -- along with the other main cosmological parameters. 
Complementary to the BAO technique, the LSS topology can also be used
as a standard ruler, by measuring and comparing the amplitude of the
genus curve at different redshifts and smoothing
lengths (see Park \& Kim 2010).
 Cosmological parameters can also be obtained
through the abundance of high-mass halos, identified as galaxy
clusters, and via several other methods that require and utilize precise knowledge of galaxy clustering 
(i.e. Zheng \& Weinberg 2007; Yoo et al. 2009). 

 
To this end, planned next generation photometric redshift galaxy surveys
will span volumes considerably larger than the current data
sets, providing a dramatic improvement in the accuracy of the
constraints on cosmological parameters.
Examples are the Panoramic Survey Telescope \& Rapid Response System
(Pan-STARRS; Kaiser et al. 2002), the Dark Energy Survey (DES; Abbott
et al. 2005), the Baryonic Oscillation Spectroscopic Survey (BOSS;
Schlegel  et al. 2009) as well as BigBOSS (Schlegel et al. 2011), the Large Synoptic Survey Telescope (LSST;
Tyson et al. 2004), the Hobby Eberly Telescope Dark Energy Experiment
(HETDEX; Hill et al. 2004), and the space-based Wide-Field Infrared
Survey Telescope (WFIRST; Green et al. 2011) and Euclid (Cimatti et al. 2008). 
The lack of precision in the redshift
determination is compensated with the larger volume of the survey, the larger density
of galaxies, and the possibility of analyzing different galaxy populations.


With large-volume surveys becoming the norm, it is timely to devise
large-volume numerical simulations able to 
mimic, reproduce and control the
various observational surveys. This paper aims at presenting
two large cosmological $N$-body simulations,
called Horizon Run 2 (HR2) and Horizon Run 3 (HR3), made
using innovative computational facilities and forefront capabilities.
Cosmological $N$-body simulations 
provide in fact a powerful tool to test the validity of cosmological
models and galaxy formation mechanisms (Bertschinger 1998), as well as they represent a
complementary and necessary benchmark to design and support the observational surveys. 
In particular, 
the advent of realistic cosmological $N$-body simulations, capable of reproducing successfully the global properties
of the observed structure at large scales, has enabled
significant progress in understanding the structure of dark
matter halos, the DM clustering over a huge range
of scales, the distribution of galaxies in space and time, the effects of the
environment, the mass assembly history, and the nature of DM itself.
For example, numerical studies have indicated that
the  hierarchical assembly of CDM halos yields approximately universal mass
profiles (Navarro et al. 1996, 1997), 
strongly triaxial shapes with a
slight preference for nearly prolate systems (i.e. Jing \& Suto 2002), presence of abundant but
non-dominant substructure within the virialized region of a halo, and
cuspy inner mass profiles (for instance see Springel et al. 2005). 
Numerical simulations are also crucial to DE studies, especially for 
characterizing BAO systematics
(Crotts et al. 2005; Crocce \& Scoccimarro 2008; Kim et al. 2009). 
These are well controlled only with a detailed 
knowledge of all nonlinear effects associated, 
and can only be addressed by studying mock catalogs constructed
from large $N$-body simulations. In this respect,
comparison with $N$-body
simulations is critical, as it allows one to calibrate the
DE experiments with high accuracy. 


While the current and future surveys of large-scale structure in the
Universe demand larger and larger simulations, simulating large
cosmological volumes with good mass resolution is clearly a major
computational challenge. However, it is crucial to be able to reach 
volumes big enough to assess the statistical significance of 
extremely massive halos, almost always underrepresented in
simulations that survey a small fraction of the Hubble volume.
For example, till recently our understanding of the
mass profile of massive haloes has been rather limited, derived
largely from small numbers of individual realizations or from
extrapolation of modes calibrated on different mass scales 
(Navarro et al. 1996, 1997; Moore et al 1998; 
Klypin et al. 2001; Diemand et al. 2004; Reed et
al. 2005). This is because, 
at high masses, enormous
simulation volumes are required in order to collect statistically
significant samples of these rare DM halos. Instead, 
individual halo simulations may result in biased
concentration estimates in a variety of ways (Gao et al. 2008).
At high masses we also 
know only approximately the redshift and mass dependence of halo
concentration, even for the current concordance
cosmology.
Another example of how important is to simulate big volumes is related to DE science:
larger volumes allow one to model more accurately the true power at
large scales and the corresponding power spectrum, particularly in the case of the LRG distribution, which
is critical to the BAO test (Kim et al. 2009).
A large box size will guarantee small statistical errors in the power spectrum estimates: we need
in fact to measure the acoustic peak scale down to an accuracy better than 1\%.

\begin{figure*}[!t]
\centering \epsfxsize=12cm 
\epsfbox{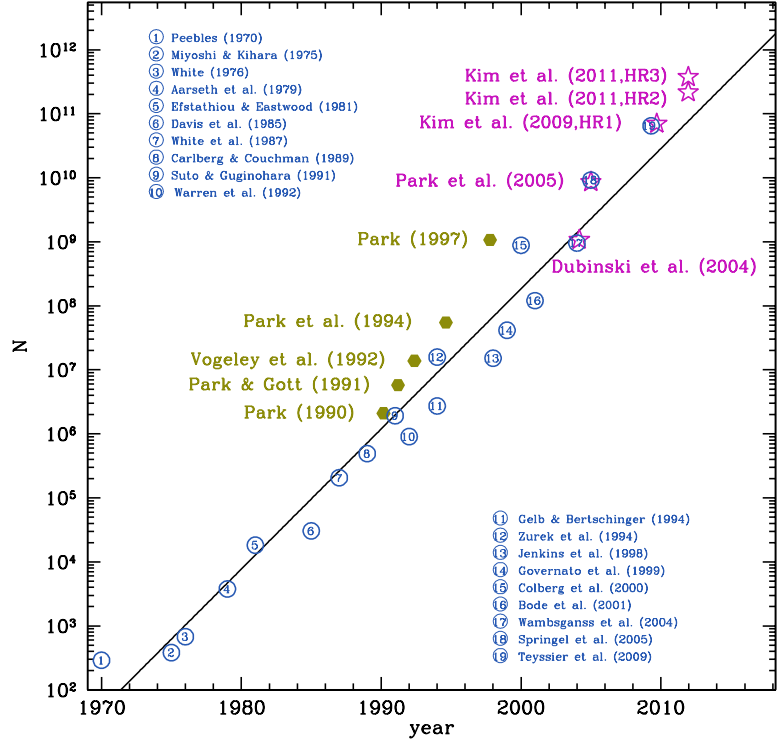} 
\caption{Evolution of the number of particles in $N$-body simulations
  versus time (years). 
Filled hexagons (pale green) are PM simulations made by our group; 
stars (pink) are PM-Tree simulations carried out by our
group (the HR2 and HR3 are located in the upper-right corner); open
  circles (blue) are those performed by other authors -- see the
  detailed legend in the figure. 
Note in particular the position of the Millennium Run (Springel et al. 2005; 10 billion particles,
  box size 500$h^{-1}$Mpc), 
and the French collaboration (Teyssier et al. 2009; 68.7 billion
  particles, box size 2000$h^{-1}$Mpc). 
The solid line is the mean evolution of
the simulation size.}
\label{fig_simulation_particles}
\end{figure*}

As explained in Neto et al. (2007) and in Gao et al. (2008), 
trying to combine different simulations of varying mass
resolution and box size is instead potentially very risky, along with
extrapolation techniques. 
To this end, Neto et al. (2007) brings the example of Macci\`{o} et
al. (2007), who combined several simulations of varying mass resolution and
box sizes in order to reach scales with $M \ll M_{*}$. Unfortunately, this
approach comes with pitfalls because in order to resolve 
a statistically significant sample of haloes with masses of the order of
$10^{10}h^{-1} $M$_{\odot}$ one must use a considerably smaller simulation box
(i.e. 14.2$h^{-1}$Mpc on a side); this implies a substantial lack
of large-scale power, due to such small periodic realization, which may
influence the result.
The obvious way out is to increase the dynamic range of the
simulation, so as to encompass a volume large enough to be
representative, while at the same time having enough mass resolution to
extend the analysis well above (or below) $M_{*}$. 

In this view, the enormous volume of our HR2 and HR3,
together with the large number
of particles used, make our simulations ideal to characterize with minimal
statistical uncertainty the dependence of the structural parameters of
CDM halos on mass, formation time, departures from
equilibrium, etc., and will be of great use for DE science and for all
the upcoming photometric redshifts surveys.


The paper is organized as follows. In Section \ref{sec_num_met} we
briefly discuss the standard numerical techniques in LSS analyses, and compare 
the number of particles and volumes of our simulations with other 
recent numerical studies. In Section
\ref{sec_our_sims}, after a
synthetic description of our previous Horizon Run 1 (HR1),
we present our new simulations HR2 and HR3. In
particular, we provide some technical details on the code used, the
mass and force resolution, and the halo selection procedure. 
In Section \ref{sec_mocks} we introduce the 35 all-sky mock surveys along the past
light cone (8 from the HR2 and 27 from the HR3), made in order to simulate the BOSS geometry. These mock
catalogs are particularly suitable for DE studies (i.e. LRGs), and
are already publicly available. We then present in 
Section \ref{sec_first_results} the 
first results from testing our new simulations; namely, the
computation of the starting redshifts, mass functions, power
spectra and two-point correlation functions.
We conclude in Section \ref{sec_conclusions}, with a brief
description on the importance and use of our simulations and catalogs.
We are making the simulation data and mocks available to the community
at \textrm{http://astro.kias.re.kr/Horizon-Run23/}.



\section{NUMERICAL METHODS}    \label{sec_num_met}

Progressively sophisticated numerical techniques are used to
carry out $N$-body simulations, 
with the goal of optimizing resources and performances by exploiting 
forefront technological developments.

Broadly speaking, there are
two basic algorithms by which a simulation can be optimized:
particle-mesh (PM) methods and tree methods -- along with
combinations of the two, such as the PM-Tree scheme.
In the PM method, space is discretised on a mesh and particles are
divided between the nearby vertices of the mesh, in order
to compute the gravitational potential via Poisson's equation and fast
Fourier transform (FFT) techniques.  
In the tree method, the volume is instead usually divided up into
cubic cells and grouped so that only particles from nearby cells need
to be treated individually, whereas particles in distant cells can be
treated as a single large particle centered at its center of
mass. Obviously, this latter scheme increases considerably the computational speed
in particle-pair interactions.

Thanks to rapid advancements in memory and
computing power capabilities, state-of-the-art simulations can be performed with
billion of particles, since
algorithmic and hardware development have increased the mass and
spatial resolution by orders of magnitude. 
Hence, it is perhaps
impressive to recall that one of the first $N$-body
simulations run by Peebles (1970) utilized only 300 particles, while 
today we can simulate individual
collapsed structures in a full cosmological context, and substructure
halos can be resolved -- with excellent approximations to CDM
halos on a large mass and spatial range.
To this end, Figure  \ref{fig_simulation_particles} shows 
how our new HR2 and HR3 simulations compare with the
previous HR1 and with other recent
large-volume simulations, in terms of number of particles; 
for specific details on our simulations refer to the next section. 
Pale green filled hexagons in the figure represent the PM simulations made by our group, 
stars (pink) are the PM-Tree simulations carried out always by our
group (see the various labels), while
blue open circles are those performed by other authors. 
Note in particular the Millennium Run (Springel et al. 2005),
with 10 billion particles and a box size of 500 $h^{-1}$Mpc, and the recent French collaboration $N$-body
simulations (Teyssier et al. 2009), made using 68.7 billion particles with
a box size of 2000 $h^{-1}$Mpc. 
The solid line is the mean evolution of
the simulation size taken from Springel et al. (2005), which is basically linear in logarithmic space,
increasing a factor of $10$ every $4.55$ years.

The application of numerical methods in cosmology is in fact an exponentially
growing field, in parallel with major technological developments;
for a brief history of numerical studies see Diemand \& Moore (2009) and
Kim et al. (2009).
So far, efforts have been essentially focused both on
investigating the formation of single halos with ultra-high resolution
(Springel et al. 2008; Stadel et al. 2009), and on simulating structure formation in
large boxes, to mimic the LSS in the
Universe. Obviously, one needs to push numerical simulations
to their limits, as the dynamical range of scales to be resolved is
extremely large for addressing different types
of cosmological problems: this study focuses on the latter part. 
What is interesting to notice here is that 
$N$-body simulations of the gravitational collapse of a collisionless
system of particles actually pre-date the CDM model (see again Diemand \&
Moore 2009). In effect,
early simulations in the 1960's studied the formation of elliptical galaxies from the collapse
of a cold top-hat perturbation of stars (van Albada 1961; Henon \& Heiles 1964;
Peebles 1970). In the 70's, attempts were made to follow the expansion of a
collapse of a spherical overdensity to relate to the observed
properties of virialized structures such as galaxy clusters.
In 1975, Groth \& Peebles made actual ``cosmological''
simulations using 1550 particles with $\Omega_{\rm M} =1$ and Poisson
initial conditions (Groth \& Peebles 1975). After, Aarseth et al. (1979) used 4000
particles. Only in the 80's, however, was it proposed that cosmic structure formation
follows a dominant, non-baryonic CDM component, and
later on the inflationary
scenario in conjunction with CDM brought realistic initial conditions for $N$-body
models. However, it was not until the simulations by Dubinski \&
Carlberg that
individual objects were simulated at sufficiently high resolution to
resolve their inner structure on scales that could be compared with
observations. Interestingly, Park (1990) was the first to use 4 million particles, a peak biasing scheme
and a CDM, $\Omega_{\rm M} h =0.2$ model to simulate a volume large enough to properly encompass the CfA Great
Wall. After that, only in 2008
the first billion particle halo simulation Via Lactea II was published
(Diemand et al. 2008), followed by Aquarius (Springel et al. 2008) and
GHALO (Stadel et al. 2009), in a progressively increment of box
size and number of particles till our HR1 (Kim et al. 2009) -- which spanned the 
largest volume at the time.

\begin{figure}[!t]
\centering \epsfxsize=7.5cm
\epsfbox{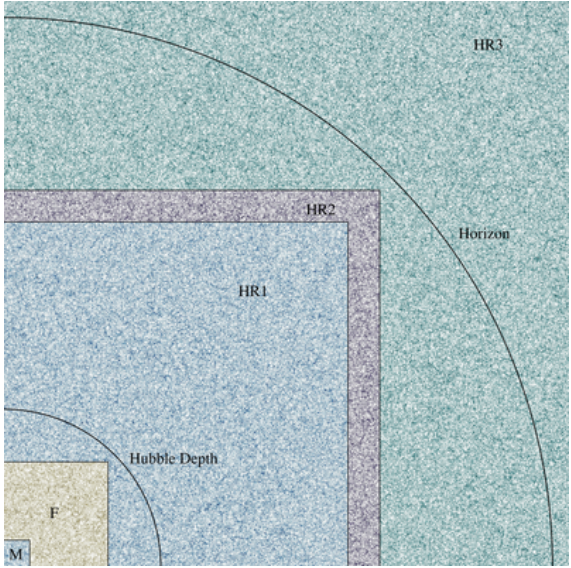} 
\caption{Visual
comparison between the box sizes of different numerical simulations,
via a 4.5 $h^{-1}$Mpc thick slice through our HR3 $N$-body
simulation  at the present epoch. 
The box sizes are 500,
2000, 6595, 7200 and 10815 $h^{-1}$Mpc for the Millennium Run,
French collaboration, HR1, HR2 and HR3, respectively. The horizon
and Hubble Depth (defined as $c H_0^{-1}$) are also indicated in the
figure, in comoving coordinates, to emphasize the
various simulation volumes.}
\label{fig_simulation_sizes}
\end{figure}

To this end, Figure \ref{fig_simulation_sizes} allows for a visual
comparison on the massive increment in box size between our Horizon Runs with other
recent cosmological $N$-body simulations. A $4.5 h^{-1}$Mpc slice
through the HR3 is performed at the present epoch, in comoving coordinates. The box sizes are 500,
2000, 6595, 7200 and 10815 $h^{-1}$Mpc for the Millennium Run,
French collaboration, HR1, HR2 and HR3, respectively. The horizon and
Hubble distance ($cH_0^{-1}$) are also indicated in the figure, to emphasize the
various simulation volumes. Note the impressive improvement with respect to the
Millennium Run box size.

\section{THE HORIZON RUNS} \label{sec_our_sims}

\begin{figure*}[!t]
\centering \epsfxsize=16cm 
\epsfbox{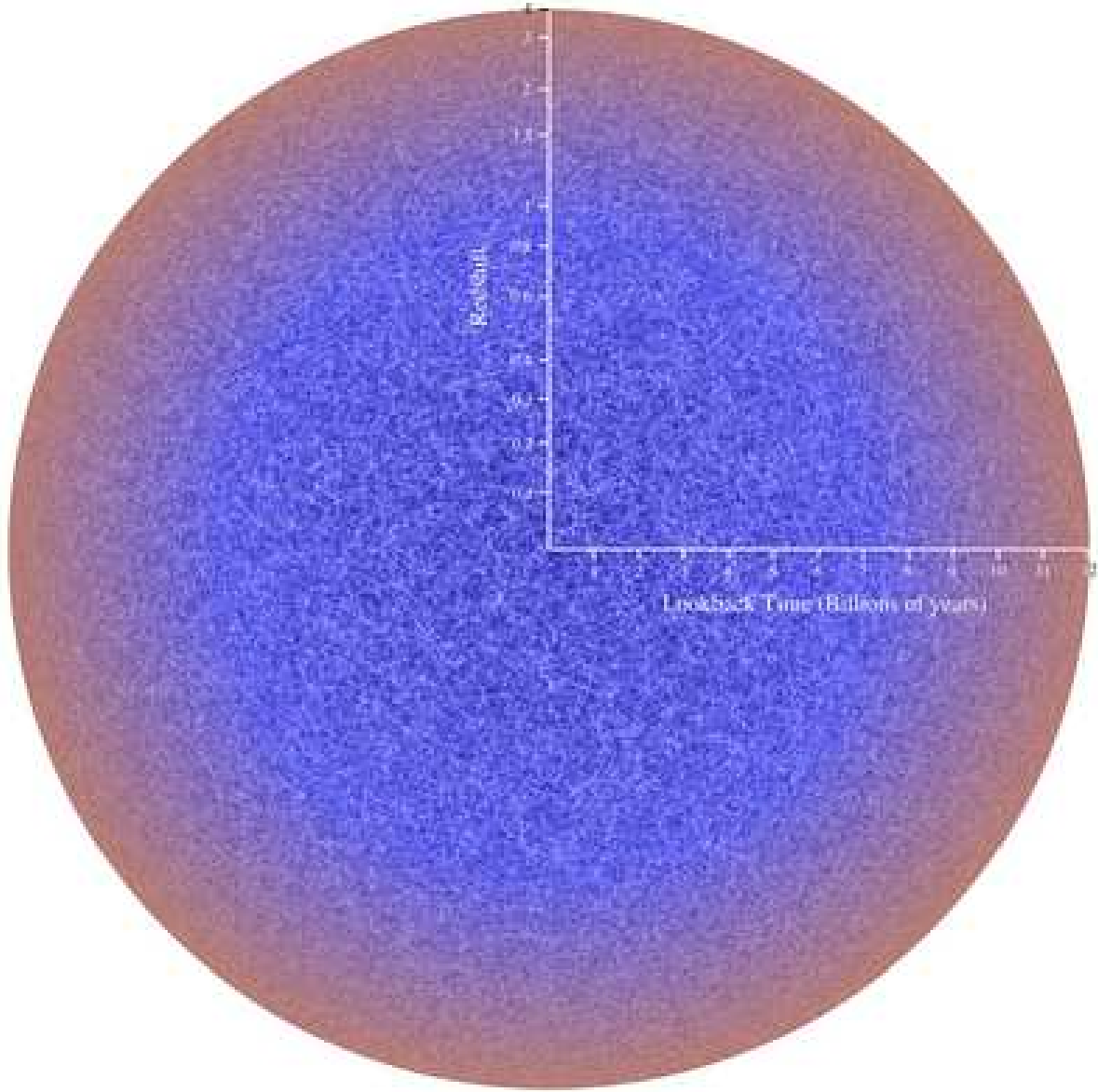} 
\caption{Example of the matter density in the past light cone space,
obtained from a slice with $30 h^{-1}$Mpc thickness 
passing through the center of the HR3 simulation box.
A mock observer is located at the center of the figure; the
radius of the map corresponds to about $z=4$. 
The ticks along the horizontal
axis indicate the lookback time, in units of billions of years, while
the redshift is indicated along the vertical axis.
The color scheme varies from blue to red as one goes to larger redshifts. 
 The radial distance is linear in lookback time. 
The starting epoch of HR3 corresponds to a redshift 
of $z=27$.}
\label{fig_simulation_light_cone}
\end{figure*}

\begin{figure*}[!t]
\centering \epsfxsize=16cm 
\epsfbox{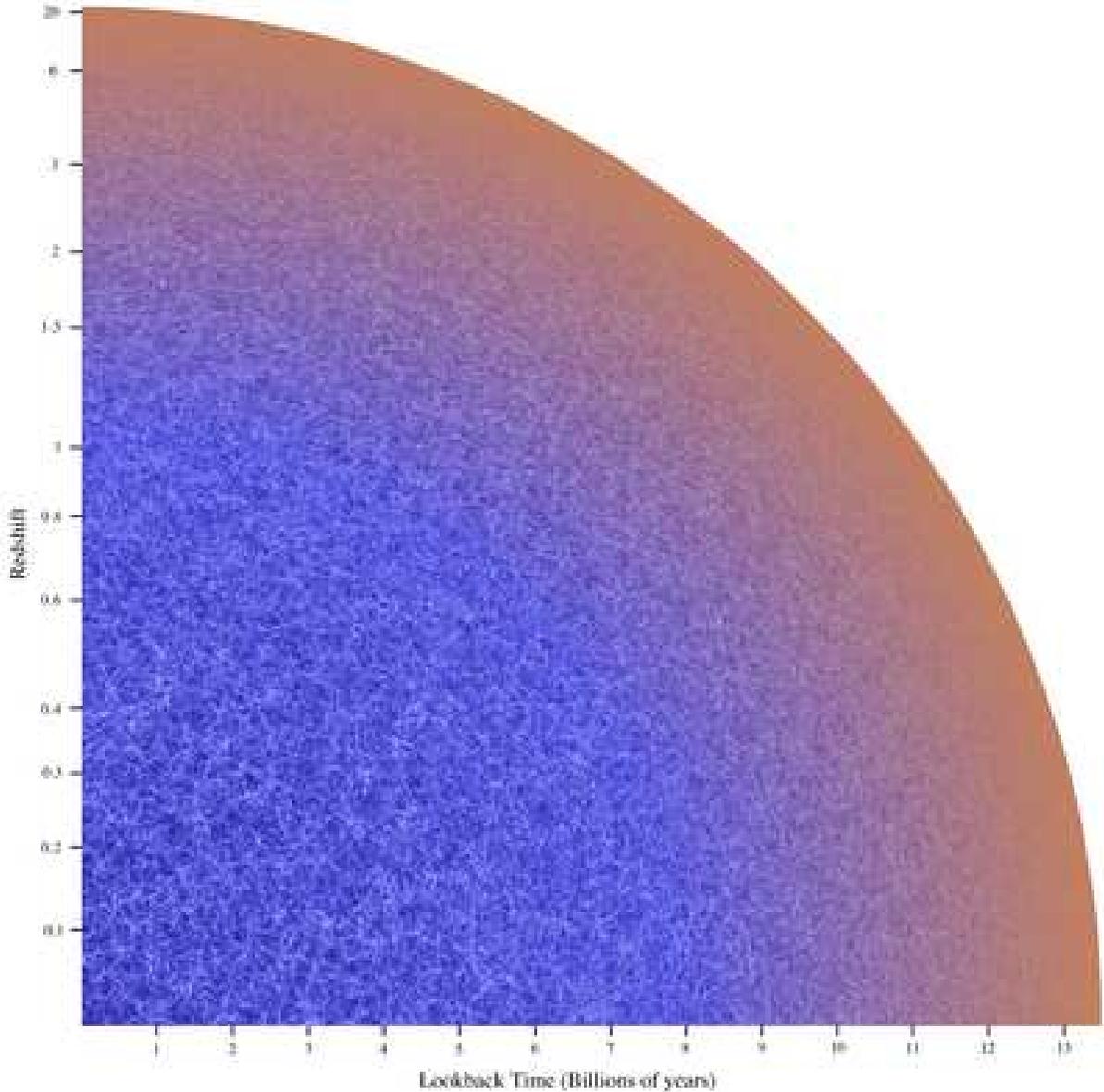} 
\caption{In false colors, a quarter-wedge slice  
density map in the past light cone is shown, from the HR3.
Color scheme as in the previous figure.
The width is $30 h^{-1}$Mpc, and the wedge reaches a redshift of $z=27$.
The density is measured using the spline kernel (a technique widely
used in SPH), with 5 nearest particles. The ticks along the horizontal axis indicate the lookback time, in
units of billions of years, while the redshift is indicated along the
vertical axis.
Note that the starting epoch of the HR3 corresponds to the maximum redshift
of $z=27$.}
\label{fig_simulation_light_cone_quarter}
\end{figure*}

\begin{figure*}[!t]
\centering \epsfxsize=15cm 
\epsfbox{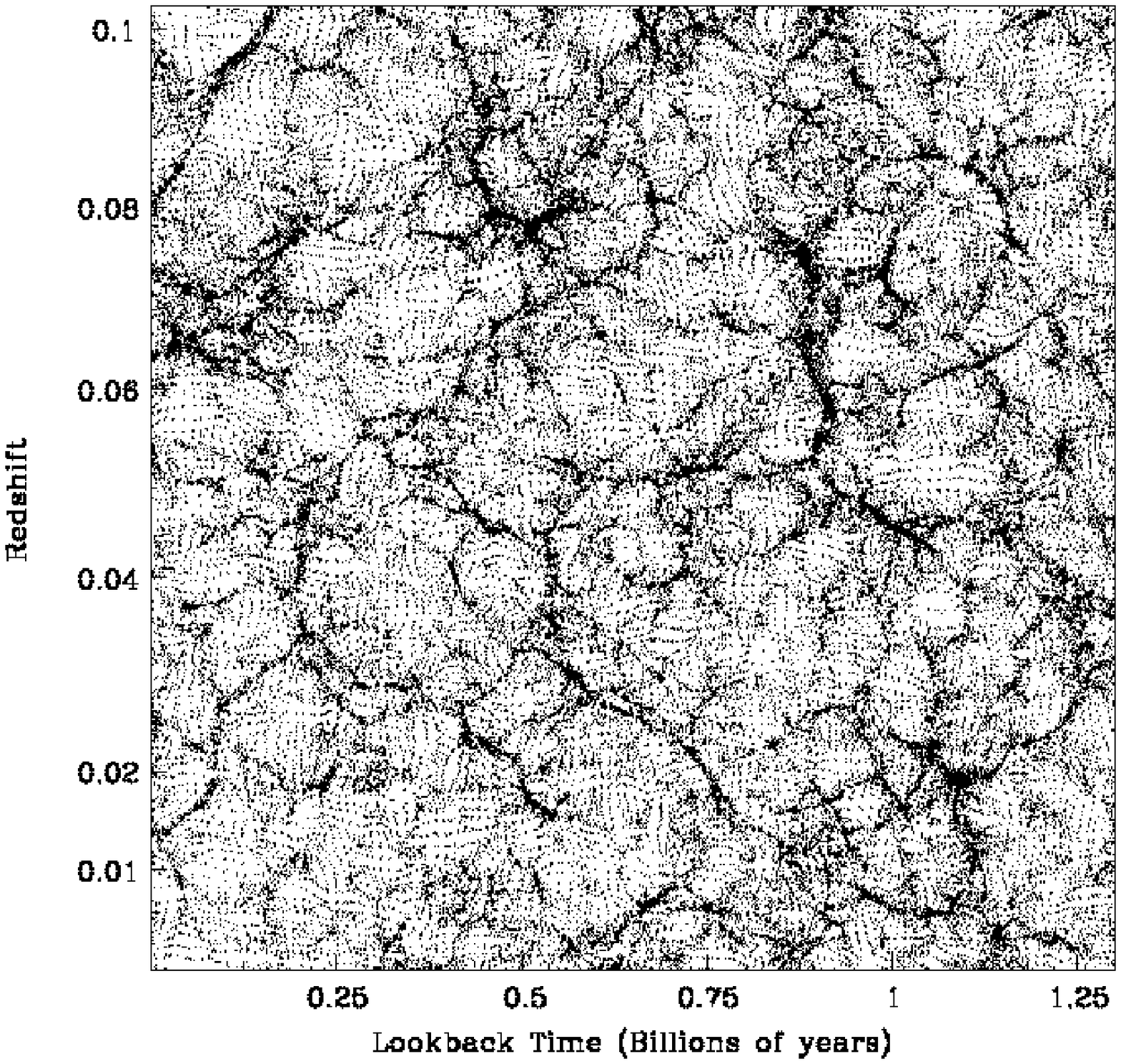} 
\caption{Particle distribution from a small region (i.e. out to $z
  \simeq 0.1$) of the entire mock survey volume
shown in Figure \ref{fig_simulation_light_cone_quarter}, which reveals 
the typical cosmic web-like pattern: 
individual halos, filaments and sheets can be easily seen, connected
in a network, with voids encompassing the volume in between. Particles
in the survey volume are shown if they are inside the slice volume with
width $\Delta z$ corresponding to 7.5 $h^{-1}$Mpc. The plot
  demonstrates the large dynamical range of our simulation, capable of
  resolving the clustering at small scales. 
}
\label{fig_simulation_light_cone_quarter_zoom}
\end{figure*}

After a brief description of our previous HR1 (Kim et al. 2009), in
this section we present the main characteristics
of the new HR2 and HR3, two of the largest cosmological $N$-body
simulations to date. 
In particular, we provide detailed technical information regarding
the code developed and used to perform the runs, the mass and force
resolution, and the halo selection procedure. The main simulation parameters are
summarized in Table \ref{table_sim_par} for convenience, and in the
following parts we will refer to it -- whenever necessary.

\begin{table*}[t]
\begin{center}
\centering
\caption{Detailed specifics of our Horizon Run $N$-body simulations.}
\doublerulesep2.0pt
\renewcommand\arraystretch{1.5}
\begin{tabular}{cccc} 
\hline \hline
 &   HR1  &   HR2 &  HR3 \\ 
\hline
Model &                               WMAP5  &      WMAP5 & WMAP5 \\
$\Omega_{\rm M}$ &                 0.26  &                         0.26             &         0.26\\
$\Omega_{\rm b}$  &               0.044   &                      0.044             &       0.044\\
$\Omega_{\rm \Lambda}$   &             0.74      &                    0.74              &        0.74\\
Spectral index  &                  0.96 &                         0.96       &                0.96\\
$H_0$ [100 km s$^{-1}$Mpc$^{-1}$]              &                        72&                             72 &                       72\\
$\sigma_8$          &                   0.794  &                     0.794    &                  0.794\\
Box size [$h^{-1}$Mpc] &             6592         &                7200        &            10815\\
No. of grids for initial conditions        &                  $4120^3$  &                   $6000^3$ &                  $7210^3$\\
No. of CDM particles  &             $4120^3$       &              $6000^3$    &              $7210^3$\\
Starting redshift    &               23         &                   32 &                        27\\
No. of global time steps  &         400             &             800   &                      600\\
Mean particle separation  [$h^{-1}$Mpc] &     1.6  &                         1.2    & 1.5 \\
Particle mass [$10^{11} h^{-1} $M$_{\odot}$]&  2.96   &                    1.25      &                   2.44\\
Minimum halo mass (30 particles) [$10^{11} h^{-1} $M$_{\odot}$] &            88.8 &                       37.5 &                         73.2\\
Mean separation of minimum mass PSB halos [$h^{-1}$Mpc] &            13.08&                       9.01&                         11.97\\
\hline
\label{table_sim_par}
\end{tabular}
\end{center}
\end{table*}


\subsection{The Horizon Run 1: overview}

The HR1 (Kim et al. 2009) was the largest volume simulation ever run
in 2009, with a box size of 6592 $h^{-1}\textrm{Mpc}$ on a side and 11418
$h^{-1}$Mpc diagonally. The cosmological model adopted was a CDM
concordance scenario with a cosmological constant (i.e. the LDCM
model), having the basic parameters fixed by the WMAP 5-year data
(Komatsu et al. 2009) listed in the second column of Table
\ref{table_sim_par}.
The initial linear power spectrum was calculated using the fitting
function provided by Eisenstein \& Hu (1998).
The initial conditions were generated on a $4120^3$ mesh with pixel
size of 1.6$h^{-1}$Mpc. It used a total of $4120^3$ = 69.9 billion CDM
particles, representing the initial density field at $z_{\rm i} = 23$.
Those particles, initially perturbed from their uniform distribution,
were gravitationally evolved by a TreePM code (Dubinski et al. 2004;
Park et al. 2005) with force resolution of 160$h^{-1}$kpc, for a total
of 400 global time steps.
The entire simulation-making process lasted 25 days on a TACHYON SUN
Blade system (i.e. a Beowulf system with 188 nodes, each with 16 CPU
cores and 32 Gigabytes of memory), at the Korean KISTI Supercomputing
Center.
The simulation used 2.4 TBytes of memory, 20 TBytes of hard disk, and
1648 CPU cores.
The entire cube data were stored at various redshifts: 
$z=0, 0.1, 0.3, 0.5, 0.7, 1$. In addition, the positions of the
simulation particles were saved in a slice of constant thickness
(equal to 64 $h^{-1}$Mpc), as they appeared in the past light cone.  
During the simulation, 8 equally-spaced (maximally-separated) 
observers were located in the cube, in order to construct the mock SDSS-III LRG surveys;  
the positions and velocities of the particles
were saved at $z<0.6$, as they crossed the past light cone.
Subhalos were then found in the past light cone data, and
used to simulate the SDSS-III LRG survey.
For more details on how the corresponding mock catalogs were
constructed see the next section, and Kim et al. (2009). 


\subsection{The Horizon Runs 2 and 3: improvements}

The major improvements of our new HR2 and HR3 with respect to the HR1
concern the number of particles used, the bigger box sizes (up to a factor of 4.4 in volume), and a
considerably finer mass resolution. 
Precisely, HR2 and HR3 have been made
using $6000^3$ = 216 billions and  $7210^3$ = 374 billion
particles, spanning a
volume of (7.200 $h^{-1}$Gpc)$^3$ and (10.815 $h^{-1}$Gpc)$^3$,
respectively -- which range from 2600 to
over 8800 times the volume of
the Millennium Run. 
The mass resolution reaches down to $1.25 \times 10^{11} h^{-1} $M$_{\odot}$, allowing to resolve galaxy-size
halos with mean particle separations of 1.2$h^{-1}$Mpc and
1.5$h^{-1}$Mpc, respectively.  Results span nearly six decades in mass.
As with the HR1, these two new simulations are based on the $\Lambda\textrm{CDM}$ cosmology, 
with parameters fixed by the WMAP 5-year data
(Komatsu et al. 2009) listed in the third and fourth columns of Table
\ref{table_sim_par}.
The initial redshifts of the simulations are $z_i=32$ and $z_i=27$,
respectively, with 800 time steps for the HR2 and 600 for the HR3.
For more details on the initial conditions, see Section \ref{subsec_ic}.
We also improved on the `old fashion' linear power spectrum by
adopting the CAMB source (http://camb.info/sources), which provides a
better measurement of the BAO scale: this is essential for a more
accurate determination of the cosmological parameters using the BAO constraint.

The positions of the
simulation particles were saved in a slice of constant thickness
(equal to 30$h^{-1}$Mpc), as they appeared in the past light cone.  
To construct the various mock catalogs, 
eight observers were evenly located in the simulation box of the HR2,
and twenty seven in the box of the HR3.
Each observer volume covers a space
spanning from $z=0$ to $z=0.7$, without overlaps: this means
that the survey volumes are totally independent.
In addition, there is one big mock survey region for each simulation.
This volume reaches $z=1.85$ (HR2) and $z=4$ (HR3), respectively, and the center is
located at the origin of the simulation box.
Note in particular that the latter mock survey covers the range out to
$z=2.5$, where BOSS data allow one to measure the 3D BAO using
Lyman-$\alpha$ clouds seen in absorption in front of many quasar lines
of sights.

Figure \ref{fig_simulation_light_cone}
is an example of the matter density in the past light cone space from
the HR3 (the width is $30 h^{-1}$Mpc). A mock observer is located at the center of the figure; the
radius of the map corresponds to about $z=4$. 
The ticks along the horizontal
axis indicate the lookback time, in units of billions of years, while
the redshift is indicated along the vertical axis. 
The color scheme varies from blue to red as one goes to larger redshifts. 
Note that the starting epoch of the HR3 corresponds to the maximum redshift
of $z=27$. The distribution of the CDM particles is converted to a density field
using the variable-size spline kernel containing 5 CDM particles -- a
technique well-known in smoothed-particle hydrodynamics (SPH).

Figure  \ref{fig_simulation_light_cone_quarter}
is another example obtained from our HR3.
In false colors, a quarter-wedge slice  
density map in the past light cone is shown.
The width is $30 h^{-1}$Mpc, and the wedge reaches $z=27$.
The color scheme varies
as in the previous figure.
The ticks along the horizontal axis indicate the lookback time, in
units of billions of years, while the redshift is indicated along the
vertical axis; the starting epoch of the HR3 is $z=27$.

Figure  \ref{fig_simulation_light_cone_quarter_zoom} shows the
particle distribution from a small region (i.e. out to $z
  \simeq 0.1$) of the entire mock survey volume, which clearly reveals
the typical cosmic web-like pattern:  one can easily distinguish  individual halos, filaments and sheets, connected
in a network, with voids encompassing the
volume in between. The plot demonstrates the large dynamical range of
our simulations, which are able to resolve correctly individual structures at
small scales. 


\subsection{Improving the GOTPM code}

The Grid-of-Oct-Trees-Particle-Mesh code, called GOTPM, is a parallel,
cosmological $N$-body code based on a hybrid scheme using the PM and
Barnes-Hut oct-tree algorithm, originally devised by Dubinski et
al. (2004). We used an improved version of this code for our HR2 and
HR3 simulations. Specifically, we implemented a new procedure which
allows us to describe more accurately the particle positions. 
In fact, when dealing with billions of particles, 
single precision (adopted in the original version of GOTPM) is inaccurate for representing their 
positions. This is because the significant bits in single
precision are only 24,
which allow for about 7 significant digit numbers.
The resulting position error in single precision is about $N/2^{24}$,
with $N$ the number of particles in one dimension. 
For example, the particles in the $7210^3$ simulation have a maximum intrinsic 
positioning error of 0.04\%, which becomes significant
when their clustering is high and when those particles are packed in small
areas. In these situations, using single precision provides highly
inaccurate results. 
A possible (and obvious) solution is to use
double precision instead; however, 
this would require too much memory space (i.e. an additional 30\% increment).
To solve the problem, we 
devised a simpler method which does not require any
additional space -- but only some more calculation time.
In essence, instead of determining the positions of the particles using the position vector in the GOTPM code, 
we used the corresponding offset vector (or displacement vector),
from the pre-initial Lagrangian position. We then calculated the 
position of the particle directly from its index and particle offset vector.
By adopting this procedure, the positioning error is about $d_{(x,y,z)}/2^{24}$, where $d_{(x,y,z)}$ is the
displacement from the Lagrangian position. This value is significantly smaller than $N$.

In addition, in the GOTPM each particle requires a total of 40 bytes; four of them
are used to save the position and velocity, eight are for the particle index, and 
other eight are allocated for a
pointer which is used to build a linked list in the Tree mode.
Since the particle pointer is not used in the PM mode, this memory
space is recycled for the density mesh and for the FFT workspace.


\subsection{Halo selection}

Identifying and deciding which material belongs to a halo and what
lies beyond it is clearly a
non-trivial question. This is because
DM halos are dynamical structures, constantly accreeting
material and often substantially out of virial equilibrium. In these
circumstances, halos evolve quickly so that the parameters used to
specify their properties change rapidly and thus are
ill-defined. Furthermore, in the case of an ongoing major merger even
the definition of the halo center becomes ambiguous. 

In the Horizon Runs, halos are first identified via a standard Friend-of-Friend (FoF)
procedure. Then subhalos are found (out of FoF halos) with a 
subhalo finding technique, developed by Kim \& Park (2006) and
Kim et al. (2008). This method allows one to  
identify physically self-bound (PSB) dark matter
subhalos not tidally disrupted by larger structures at the desired
epoch. 
This procedure 
does not
discard any information on the subhalos which is actually in the $N$-body
simulations, and would be thrown away in a Halo Occupation Distribution
(HOD) analysis using just simple FoF halos.
Note in fact that 
the FoF algorithm is a percolation scheme that makes no assumptions
about halo geometry, but may spuriously group distinct halos together
into the same object.
In particular, we have applied the parallel version of the FoF halo finding method
to identify virialized structures from the simulation particles.
Linked particles with mutual distance less than $0.2 \times d_{\rm mean}$,
with $d_{\rm mean}$ the mean particle 
separation, are grouped. In addition, 
subhalos in a FoF halo are found by a new version of the PSB method, 
more advanced than the previous one (Kim \& Park 2006) in that it adopts
adaptive density fields rather than the rectangular grid density fields.
Densities at the particle positions are measured using the SPH density
allocation scheme, and each particle is linked to 30 nearest neighbors.
The neighbor links are used for building hierarchical isodensity 
contours around local density peaks.
These coordinate-free density allocation and neighbor search are much 
more effective for identifying
subhalos in crowded regions.



\section{MOCK SURVEYS} \label{sec_mocks}

In this section, first we briefly describe
our previous mock catalogs made from the HR1, in support of the SDSS-III.
We then introduce our new 
35 all-sky mock surveys along the past
light cones, made  from the HR2 and HR3 to simulate the BOSS geometry and 
already publicly available at \textrm{http://astro.kias.re.kr/Horizon-Run23/}.

Regardless of the specific simulation considered, the following general procedure is used to construct 
a mock catalog. 
For each simulation time step, we track particles
which are located in the light cone shell surrounded by two spherical surfaces
of radii $(d-\Delta_{1/2},d+\Delta_{1/2})$ (i.e. inner and outer shells), with 
$d$ being the comoving distance from the observer at the given time step,
and $\Delta$ the difference between the comoving distance 
at the next step (subscript 1) or at the previous step (subscript 2) --
with respect to that at the current step.
We then calculate the particle
distances and flag them if they are in the shell. 
The radius and width of the shell (and so the volume) are
calculated according to the cosmological model and
the simulation time and step size. Obviously, 
the particles can cross the shell boundary between different
time steps, since they are moving in space. If that happens, 
they can be double
detected in adjacent time steps -- or totally missed.
In the former case, if a particle is double detected between consecutive time steps,
we average its position and velocity.
For the latter case, in order not to miss a particle, we create a buffer zone with constant
width ($\Delta_{\rm R}=0.1$ of the simulation pixel), and make two regions
in the inner and outer shell boundaries, respectively; 
we then identify particles inside the buffer zone. 
If a particle is detected in both regions of the buffer zone,
we average its position and velocity
and save the information in the light cone particle list.
This method has been devised to account for double detections, 
and in order not to miss
particles -- which happens frequently due to the finite size of the
simulation time step.


\subsection{Mocks from the HR1: overview}

To make mock SDSS-III LRG surveys from the HR1, we placed observers at
8 different locations in the simulation cube, and carried out all-sky surveys 
up to $z=0.6$. These are all past light cone data. 
The effects of our choices of the starting redshift, time step and
force resolution were discussed in detail in Kim et al. (2009).
In particular, we used the Zel'dovich
redshifts and the FoF halo multiplicity function to estimate their
effects (see also the next section).

During the simulation we located 8 equally-spaced (maximally-separated) 
observers in the simulation cube, and saved the positions and
velocities of the particles
at $z<0.6$ as they cross the past light cone.
We assumed that the SDSS-III survey would produce a volume-limited LRG sample
with constant number density of $3\times 10^{-4} (h^{-1}\textrm{Mpc})^{-3}$.
In our simulation we varied the minimum mass limit of subhalos to match
the number density of selected subhalos (the mock LRGs) with this number
at each redshift.
For example, the mass limit yielding the LRG number density of 
$3\times 10^{-4} (h^{-1}{\textrm{Mpc}})^{-3}$ 
was found to be $1.33\times10^{13} h^{-1}$M$_{\odot}$
and $9.75\times10^{12} h^{-1}$M$_{\odot}$ at $z=0$ and $z=0.6$, respectively.
We then checked how well the mock
LRG sample reproduced the physical properties of the existing LRG
sample -- see for this Kim et al. (2008), Gott et al. (2008), Gott et
al. (2009), Choi et al. (2010). We found a good match from 1 to 140$h^{-1}$Mpc, particularly in
the case of the shape of the correlation function. 
In addition, the BAO scale and the LSS topology could be very
accurately calibrated with these mocks. 


\subsection{New mocks from the HR2 and HR3}

\begin{figure}[!t]
\centering \epsfxsize=8.5cm 
\epsfbox{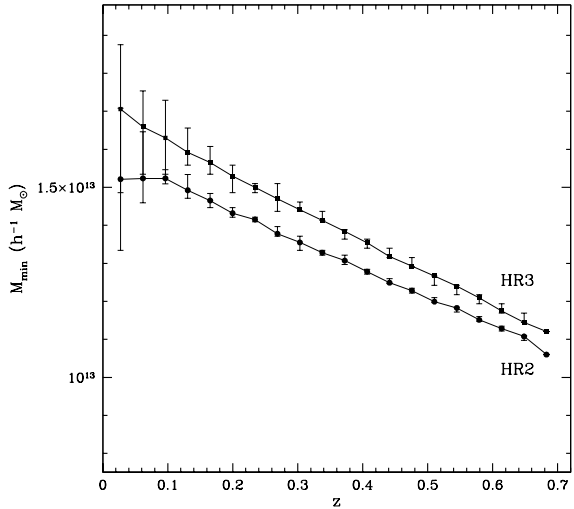} 
\caption{Minimum PSB halo mass as a function
of redshift for the HR2 and HR3, constrained to give the mean halo number density 
of $3 \times 10^{-4}(h^{-1}\rm{Mpc})^{-3}$. This plot is useful 
for constructing mock catalogs which simulate the distribution of LRGs
in the BOSS survey. See the main text for more details.}
\label{fig_psb}
\end{figure}

To simulate the ongoing BOSS survey and construct the various mock catalogs,
we  evenly  positioned 
observers at the centers of eight (HR2) and twenty seven (HR3)
sub-cubes of the entire simulation boxes.
Each observer volume covers a space
spanned from $z=0$ to $z=0.7$, without overlaps: this means
that the survey volumes are totally independent.
In addition, we considered one mock observer covering the entire simulation region, 
with survey depth down to $z=1.85$ for the HR2, and $z=4$ for the HR3.
For each simulation time step, we track particles
which are located in the light cone shell surrounded by two spherical
surfaces as explained before, and 
account for double detections and potentially missing particles.

In particular, our mocks can be used to
simulate the distribution of LRGs
in the BOSS survey. 
This is done as follows: first, obtain FoF halos
with standard linking length of $0.2 \times d_{\rm mean}$ from 
the mock particle distributions. 
Then identify subhalos in each FoF halo using the PSB method (Kim et al. 2008),
and call them PSB halos.
Kim, Park, \& Choi (2008) showed  that there is a strong
one-to-one relation between PSB halos and galaxies,
under the condition that a more massive PSB halo contains a brighter
galaxy. Hence, apply this relation to the PSB halo
list to simulate the distributions of LRGs -- note that their mean observed number density
is expected to be $\simeq 3 \times 10^{-4} (h^{-1}\textrm{Mpc})^{-3}$.
It is then possible to find the minimum mass of PSB halos which can have
LRGs with luminosity higher than the BOSS observation limit, by
using their expected number density.
In Figure \ref{fig_psb} we show the minimum PSB halo mass as a function
of redshift for the HR2 and HR3, constrained to give the mean halo number density of $3 \times 10^{-4}(h^{-1}\rm{Mpc})^{-3}$ --
note that the  minimum PSB halo mass is a fixed number in each simulation.
In the figure, the two trends are similar, independently of the
simulation considered, but their amplitudes differ
substantially. This is mainly due to the different resolutions of the simulations.
Clearly, a finer resolution allows us to detect subhalos
which are not seen at lower resolutions. 
Consequently, massive subhalos can be divided into several
less massive subhalos at higher resolutions --
the well-known cloud-in-cloud problem.
This is also why in Figure \ref{fig_psb} the HR2 results have lower
amplitude, since the HR2 has the highest mass resolution with respect
to the HR3. Therefore, for a given mass scale,
the HR3 has more subhalos than HR2 and a larger minimum
PSB halo mass than HR2. 



\section{FIRST RESULTS} \label{sec_first_results}

In this section we present the first results from the analysis of the
HR2 and HR3 simulations. In particular, 
we have measured power spectra, correlation functions, mass
functions and basic halo properties with percent
level accuracy, and verified that they correctly reproduce the $\Lambda\textrm{CDM}$
theoretical expectations, in excellent agreement with linear
perturbation theory. In what follows, we also provide some more
details on how the initial conditions have been fixed.
The results described here are only basic checks, as the main goal of
this study is to present our new simulations;
in the upcoming papers we will provide more detailed scientific
analyses, with a particular focus on topology, BAO,
and galaxy formation science.


\subsection{Initial conditions: starting redshift and time steps} \label{subsec_ic}

\begin{figure}[!t]
\centering \epsfxsize=8.75cm 
\epsfbox{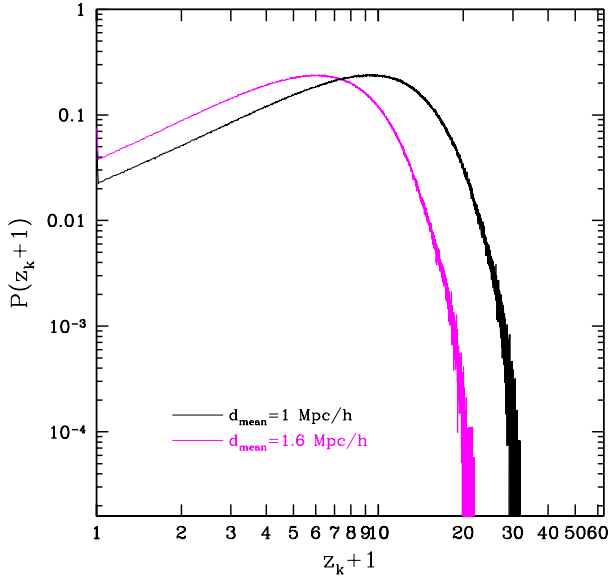} 
\caption{Distributions of the Zel'dovich redshifts
for two different pixel sizes  (see the main text for definitions). 
 The black curve corresponds to $
  d_{\rm mean}=1 h^{-1}$Mpc, while for the purple one $d_{\rm mean}=1.5 h^{-1}$Mpc.
This figure is used to determine the starting redshifts of the HR2
and HR3 simulations.}
\label{fig_zeld}
\end{figure}

We used a TreePM code (Dubinski et al. 2004) to generate the initial Zel'dovich
displacements for our HR2 and HR3, which adopts the first-order
Lagrangian perturbation scheme.
Correctly determining the proper initial conditions is essential for $N$-body simulations.
In fact, as 
Crocce et al. (2006) pointed out, improper initial conditions may cause
errors in the resulting halo mass function and bias function for massive halos. Those errors originate by
starting the simulations at too low redshift, while using first-order
techniques for the initial particle displacements and velocities. 
Obviously, a way out is to
use second order perturbation theory to correct for these effects.
Alternatively, one can simply adopt the   first-order
Lagrangian perturbation scheme and choose the 
starting redshifts of the simulations
so that the Lagrangian shifts of the particles 
are not larger than the pixel spacing: this correspond to $z_{\rm i}=32$ for the HR2, and  
$z_{\rm i}=27$ for the HR3 (see the discussion below).
In fact, if the initial displacement is larger than the pixel spacing,
a particle may not fully experience the local differential gravitational potential,
and consequently the shift calculated using the first-order scheme
could not fully reflect the fluctuations at the pixel scale.
Figure \ref{fig_zeld} shows how we quantitatively estimate the proper
starting redshifts for our initial conditions. 
As in Kim et al. (2009), we introduce the Zel'dovich
redshift $z_{\rm k}$ of the particles, defined as 
the epoch when its
Zel'dovich displacement becomes equal to the pixel size -- either in
the $x$, $y$ or $z$ directions.
Two simulations are considered, 
tagged by the mean particle 
separation $d_{\rm mean}$, or equivalently by the pixel size.
In the first one  $d_{\rm mean}=1 h^{-1}$Mpc -- a value close to that of the HR2, which has
a mean separation of 1.2$h^{-1}$Mpc; in the second one $d_{\rm mean}=1.6
h^{-1}$Mpc, and is used for the HR3, whose 
mean separation is 1.5$h^{-1}$Mpc.
To measure the distributions,
we perform an initial setting moving $256^3$ particles from the initial
conditions defined on a $256^3$-size mesh.
Each distribution shows a power-law increase with $z_{\rm k}$
and a sharp drop after a peak.
It can be clearly seen that no particle experiences a shift larger than the pixel size at $z=32$
when $d_{\rm mean}=1h^{-1}$Mpc (black curve, at the HR2 resolution), and
at $z=27$ when $d_{\rm mean}=1.5h^{-1}$Mpc (purple curve, at the HR3
resolution). This justifies our safe choice of the initial redshifts for
the HR2 and HR3 simulations.

We adopted 800 time steps to evolve the CDM particles from $z=32$ to 0 for
the HR2, 600 time steps from $z=27$ to 0 for the HR3, and set the
force resolution to 0.1$ \times d_{\rm mean}$. 
The choice of the various time steps is motivated next.


\subsection{Mass functions}

\begin{figure}[!t]
\centering \epsfxsize=8.0cm 
\epsfbox{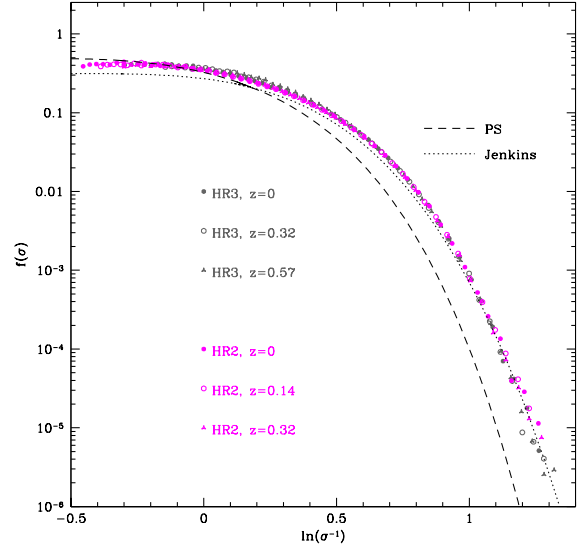} 
\caption{Halo multiplicity function computed from the HR2 and HR3. A
   standard FoF identification scheme is adopted. 
   The HR2 results are displayed in purple, those of the HR3 in grey.
Filled circles are measurements performed at 
$z=0$, open circles correspond to a redshift $z=0.14$ for the HR2 and
to $z=0.32$ for the HR3, and
filled triangles are at $z=0.32$ for the HR2 and at $z=0.57$ for the HR3. 
The minimum halo mass which
can be resolved in the HR2 (30 simulation particles)
corresponds to $M_{30} \simeq 3.75 \times 10^{12} h^{-1} $M$_{\odot}$,
while $M_{30} \simeq 7.32 \times 10^{12} h^{-1} $M$_{\odot}$ for the HR3.
The dashed line shows the Press \& Schechter (1974) mass
function (PS); the dotted line is instead the fit provided by Jenkins
   et al. (2001). See the main text for more details.}
\label{fig_mf}
\end{figure}

Because of the relatively large step size and the lower force resolution,
the question arises as to whether or not
the HR2 and HR3 simulations
have sufficient time and force resolution to correctly model 
the formation of halos.
To this end, we can simply evaluate if the 
simulations retain sufficient power
to resolve small structures by 
comparing
the FoF halo multiplicity functions computed directly from the 
HR2 and HR3 with numerical fitting formulae 
obtained from high-resolution simulations.
This is particularly true for 
the population of less massive halos:
in fact, the  coarse time step size or a poor force resolution may
destroy smaller structures more easily.
By computing the halo multiplicity functions, Figure \ref{fig_mf} shows 
that indeed our large-volume simulations are able to resolve correctly 
about six decades in mass. 
Recall that the multiplicity function, a quantity independent of
redshift, is the differential distribution function of the normalized
fluctuation amplitude of dark halos for each mass element, defined as 
\begin{equation}
f[\sigma(M)] = n(M) \Big | {{\rm d}M   \over {\rm d} {\rm ln} \sigma }
\Big | { M \over \bar{\rho}}
\end{equation}
where $n(M)$ is the number density of halos with mass $M$,
$\sigma(M)$ is the standard deviation of the density field smoothed with
a tophat of mass scale $M$, and $\bar{\rho}$ is the background density. 

In the figure, the HR2 results are displayed in purple, those of the HR3 in grey.
Filled circles are measurements performed at 
$z=0$, open circles correspond to a redshift $z=0.14$ for the HR2 and
to $z=0.32$ for the HR3, and
filled triangles are at $z=0.32$ for the HR2 and at $z=0.57$ for the HR3. 
In particular, the minimum halo mass which
can be resolved in the HR2 (30 simulation particles are used)
corresponds to $M_{30} \simeq 3.75 \times 10^{12} h^{-1} $M$_{\odot}$,
while $M_{30} \simeq 7.32 \times 10^{12} h^{-1} $M$_{\odot}$ for the HR3.
In the plot, the dashed line shows the Press \& Schechter (1974) mass
function (PS); the dotted line is instead the fit provided by Jenkins et al. (2001).
From the figure it can be seen that the Jenkins et al. (2001) curve
provides a good fit to our measurements, with only minor deviations
(but see Lukic et al. 2007 for more details on this scatter).

Our measurements at different redshifts are all 
consistent. Hence, we safely conclude that 
the FoF halo multiplicity function is complete
down to the simulation mass resolution for halos: the
simulations correctly model the formation of halos, and are
not affected by our choices of the force resolution and the number of 
time steps.


\subsection{Power spectra and correlation functions}

\begin{figure*}[!t]
\centering \epsfxsize=16cm 
\epsfbox{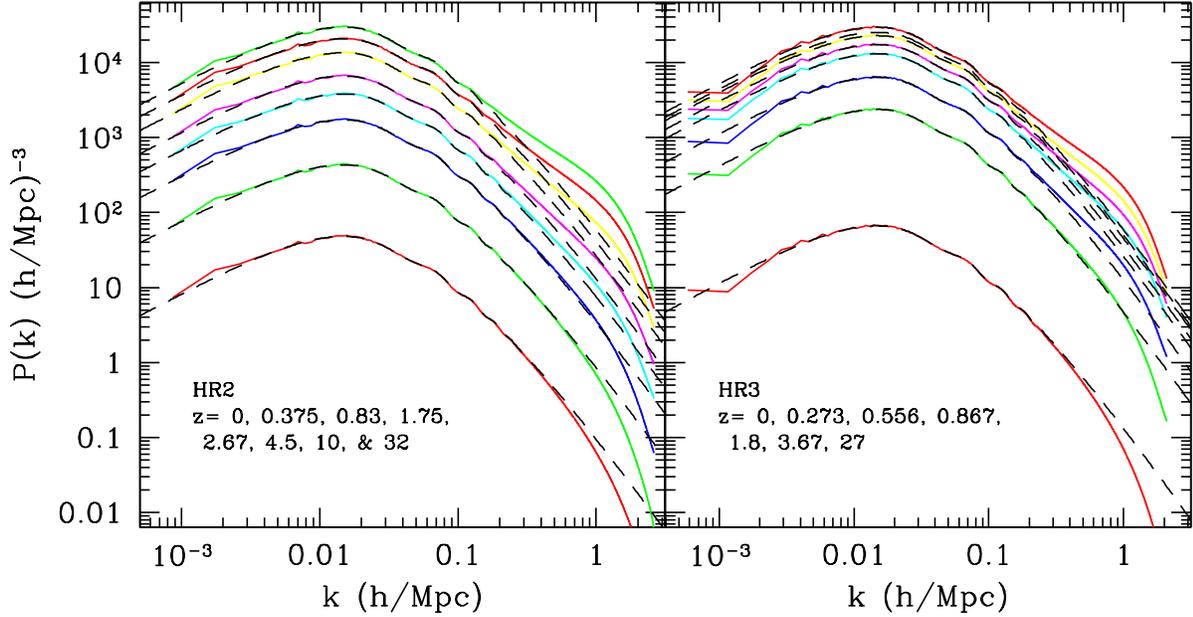} 
\caption{Power spectra of the CDM density field at various redshifts
  in comoving coordinates,
  as measured from the HR2 (left panel) and HR3 (right panel). The
  lower lines in the figure are the initial
  conditions, and the top ones are at $z=0$. Solid lines are measurements
  from simulations; long-dashed lines are obtained by linearly
  evolving the (linear) power spectra to the corresponding
  redshifts.}
\label{fig_power_spectra}
\end{figure*}

\begin{figure*}[!t]
\centering \epsfxsize=16cm 
\epsfbox{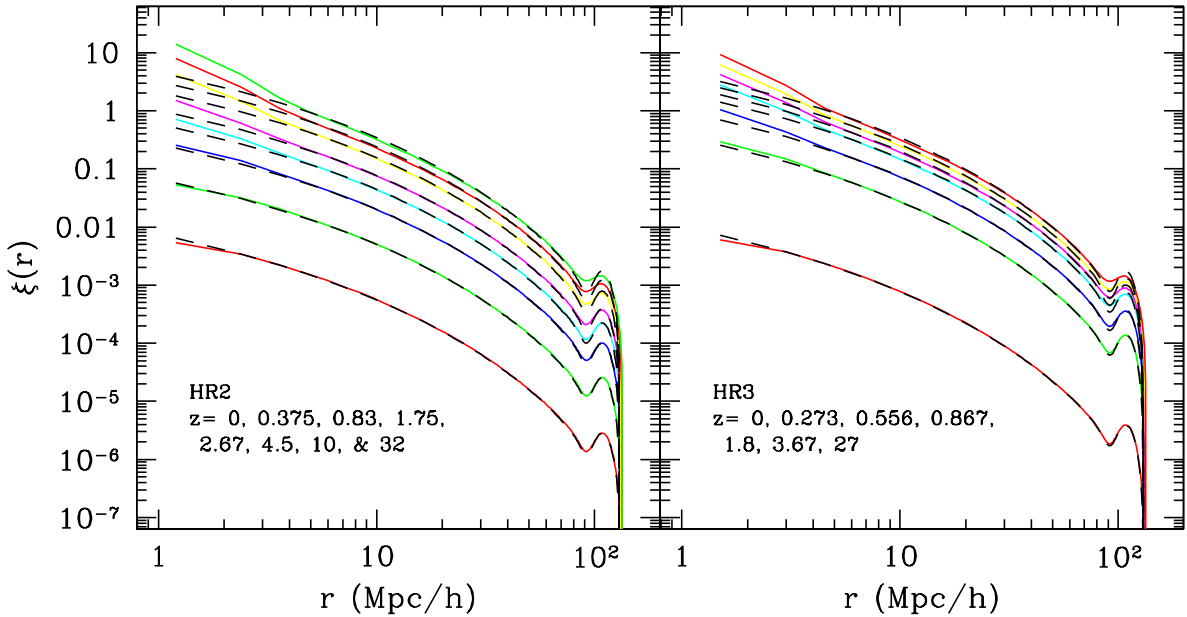} 
\caption{Two-point correlation functions of the CDM density field at
  various redshifts in comoving coordinates,
  as measured from the HR2 (left panel) and HR3 (right panel). As in
  the previous figure, the
  lower lines are initial
  conditions and the top ones are at $z=0$. Solid lines are again measurements
  from simulations; long-dashed lines are obtained by Fourier
  transforming the linearly
  evolved (linear) power spectra to the corresponding
  redshifts. Note the BAO bump, at different redshift intervals.}
\label{fig_cf}
\end{figure*}

Next, we compute the power spectrum and two-point correlation
function for each individual large-volume simulation.  
In particular, Figure \ref{fig_power_spectra} shows the  
power spectra of the CDM density field at various redshifts in
comoving coordinates, 
as measured from the HR2 (left panel) and HR3 (right panel). 
Solid lines in the figure are simulation results, while long-dashed lines
are obtained by linearly evolving  
the (linear) power spectra to the corresponding redshifts.
Note that the redshift increases from the bottom to the top of the
panels; therefore the lower curves correspond to the initial
conditions (i.e. $z=32$ for the HR2, and $z=27$ for the HR3), while
the top ones are at $z=0$. 

Figure \ref{fig_cf} shows the corresponding evolution of the two-point correlation
function of the matter density field (in comoving coordinates), for the same
redshift intervals considered in the previous figure. 
Again, the left panel highlights results from the HR2, while 
the right one presents measurements from the HR3.  
Solid lines are simulation results, long-dashed lines
are obtained by Fourier transforming the linearly evolved 
(linear) power spectra to the corresponding redshifts.
Note that the lower curves correspond to the initial
conditions, while
the top ones are at $z=0$. The BAO bumps at different redshifts are 
clearly visible; in a dedicated forthcoming paper, we will present a detailed study on the statistical
significance of the BAO feature
in the HR2 and HR3. Note the tiny amount of noise because we are
sampling such a large volume, which allows us to reduce considerably the
uncertainty in the BAO bump position.
In addition, the systematics on the BAO position due to nonlinear
effects and biasing are small, and can be corrected for with high
accuracy. 



\section{CONCLUSIONS} \label{sec_conclusions}


The main goal of this paper was to present in detail
two new large cosmological $N$-body simulations, HR2 and HR3, made
using $6000^3$ = 216 billions and  $7210^3$ = 374 billion
particles, and spanning a
volume of (7.200 $h^{-1}$Gpc)$^3$ and (10.815 $h^{-1}$Gpc)$^3$,
respectively. 
The volumes of these two simulations are considerably bigger than our
previous HR1 volume,
up to a factor of 4.4, and the mass resolution has been improved (down
to $1.25 \times 10^{11} h^{-1} $M$_{\odot}$);
this allows to resolve large-galaxy (i.e. LRG galaxy) sized  
halos, with mean particle separations of 1.2$h^{-1}$Mpc (HR2) and
1.5$h^{-1}$Mpc (HR3). The main characteristics of the
simulations have been described in detail in Section
\ref{sec_our_sims}, including 
the code developed and used to perform the runs, the mass and force
resolution, and the halo selection procedure. The basic simulation parameters are
listed in Table \ref{table_sim_par}.

Planned next generation photometric redshift galaxy surveys such
as BOSS (Schlegel  et al. 2009), BigBOSS (Schlegel et al. 2011), 
DES (Abbott et al. 2005), LSST (Tyson
et al. 2004), WFIRST (Green et al. 2011) and Euclid (Cimatti et al. 2008)
will span volumes considerably larger than the current data
sets, providing a dramatic improvement in the accuracy of the
constraints on cosmological parameters; to this end, our simulations will be of
great help for these surveys.

In particular, in support of the SDSS-III, we made a total of 35 all-sky mock catalogs along the past
light cone up to $z=0.7$, 8 from the HR2 and 27 from the HR3, to simulate the BOSS
geometry (Section \ref{sec_mocks}).
The simulations and mock surveys are already publicly available to the
community, and they should be useful for a variety of
studies in cosmology and astrophysics, ranging from large-scale structure
topology, baryon acoustic oscillations, dark energy and the
characterization of the expansion history of the
Universe, till galaxy formation science.

In Section \ref{sec_first_results} 
we also presented the first results from the analysis of the
HR2 and HR3 simulations. In particular, 
we have measured power spectra, correlation functions, mass
functions and basic halo properties with percent
level accuracy, and verified that they correctly reproduce the $\Lambda\textrm{CDM}$
theoretical expectations, in excellent agreement with linear
perturbation theory. For example,
by computing the halo multiplicity functions we proved
that our large-volume simulations are able to resolve correctly 
about six decades in mass. Much more scientific analyses can be carried out
with these simulations, and in
the upcoming papers we will focus in particular on topology, BAO,
and galaxy formation science.


Clearly, one may question the utility of performing such large-volume, expensive and
computationally challenging simulations. Why not running instead many smaller
simulations, with different initial conditions?   
Several reasons were already pointed out in the introductory part;
essentially, the main goal in doing so is to cover large scale modes
accurately, which is crucial for dark energy and BAO science
(i.e. systematic effects), for the statistical study of rare
massive halos, and for the formation of unusually large scale
structures. 
In fact, as we argued at the beginning, 
extremely massive halos are almost always underrepresented in
simulations that survey a small fraction of the Hubble volume, and 
in order to reach a statistically significant sample of these objects
one needs simulations with enormous dynamic range.  
By contrast, combining different simulations of varying mass
resolution is instead potentially very risky (i.e. Neto et al. 2007; Gao et
al. 2008). 

While we strive to carry out
simulations in as large volumes as possible, we also need high resolutions 
to sample accurately the underlying gravitational potential wells.
The HR2 and HR3 achieve both criteria: their volumes are unprecedentedly
large, but their mass resolutions
allow us to resolve large-galaxy-sized (LRG) halos; therefore,  
they are ideal to characterize with minimal
statistical uncertainty the dependence of the key structural parameters of
CDM halos on mass, spin, formation time, etc.,
and will be of great use for DE science (i.e. BOSS) and for all
the upcoming photometric redshifts surveys.

In particular, a topical issue which can be addressed with our
simulations is the nature of DE and the determination of the 
DE EoS parameter $w(z)$ as a function of redshift. A standard
technique, which has received much attention lately, is the 
measurement of the BAO scale
using LRGs (Eisenstein et al. 2005). In this case, being able to 
control the systematics is imperative, and can only be achieved via
large volume $N$-body simulations such as the HR2 and HR3.
Note in fact that the typical box sizes usually
adopted for these studies in previous literature range from
500$h^{-1}$Mpc to 2000$h^{-1}$Mpc, but those boxes are
inadequate because they are too close to the 
 baryon oscillation scale -- which is of the
order of 108 $h^{-1}$Mpc. Instead, it is essential to be able to model the power
spectrum accurately at large scales, and to have a large box size so
that the statistical errors in the power spectrum are small.

The HR2 and HR3 also permit us 
to assess the statistical significance
of unusually large structures such as the Sloan Great Wall which, with a length of 1.37
billion light years (Gott et at. 2005),  
is considerably more massive than predicted by previous $N$-body
simulations (i.e. Springel et al. 2005). 
The large number of halos in our simulations allows us to study
in detail deviations from the mean trends and the possible presence of
such systems with unusually high concentrations of galaxy haloes, or
unusually low densities. 
We will present a detailed study focused on this issue in a separate publication. 
We will also investigate several other applications,   
which involve the large scale topology in relation to dark energy and distance
measurements, the detailed modeling of the integrated Sachs-Wolf effect
(Yoo, Fitzpatrick, \& Zaldarriaga 2009) and the comparison
with counts of LRGs in the sky, 
and even applications of first
order corrections to our simulations in order to capture general relativity effects.


Our simulations and mock surveys are already publicly available at 
\textrm{http://astro.kias.re.kr/Horizon-Run23/}.
Please cite this paper whenever you use the simulation data.



\acknowledgments{
This work was supported by the Supercomputing Center/Korea Institute
of Science and Technology Information with supercomputing resources,
including technical support (KSC--2011--G2--02). We thank the Korea
Institute for Advanced Study for providing computing resources (KIAS
Center for Advanced Computation Linux Cluster System) for this work.}



{}


\end{document}